\documentclass[%
 reprint,
 amsmath,amssymb,
 aps,
]{revtex4-2}

\usepackage{graphicx}
\usepackage{dcolumn}
\usepackage{bm}
\usepackage{hyperref}
\usepackage{xcolor}
\usepackage{array}
\usepackage{enumitem}

\newcommand{\comment}[1]{}
\newcommand{\D}{\ensuremath{\mathrm{d}}}

\begin{document}

\preprint{APS/123-QED}

\title{Dynamics of non-Markovian systems:\\ Markovian embedding vs effective mass approach}

\author{Mateusz Wi\'{s}niewski}
\author{Jakub Spiechowicz}%
\email{jakub.spiechowicz@us.edu.pl}
\affiliation{%
 Institute of Physics, University of Silesia, 41-500 Chorz\'{o}w, Poland
}%


\begin{abstract}
Dynamics of non-Markovian systems is a classic problem yet it attracts an everlasting activity in physics and beyond. A powerful tool for modeling such setups is the Generalized Langevin Equation, however, its analysis typically poses a major challenge even for numerical means. For this reason, various approximations have been proposed over the years that simplify the original model. In this work we compare two methods allowing to tackle this great challenge: (i) the well-known and successful Markovian embedding technique and (ii) the recently developed effective mass approach. We discuss their scope of applicability, numerical accuracy, and computational efficiency. In doing so we consider a paradigmatic model of a free Brownian particle subjected to power-law correlated thermal noise. We show that when the memory time is short, the effective mass approach offers satisfying precision and typically is much faster than the Markovian embedding. Moreover, the concept of effective mass can be used to find optimal parameters allowing to reach supreme accuracy and minimal computational cost within the embedding. Our work therefore provides a blueprint for investigating the dynamics of non-Markovian systems.
\end{abstract}

\maketitle


\section{Introduction} \label{sec:introduction}

The system dynamics is non-Markovian if it depends not only on the current but also on its past states, or, in other words, if the system exhibits memory.
There are numerous examples of such setups in nature, including particles in viscoelastic media \cite{Banerjee2022, Ginot2022, Goychuk2022, Ferrer2021, Narinder2018, Gomez2016, Goychuk2012}, proteins \cite{deSancho2014, Ayaz2021}, spin glasses \cite{Baity-Jesi2023}, and active matter \cite{Kanazawa2020, Militaru2021, Tucci2022, Cao2023}, to name just a few.
The origin of memory typically lies in a high complexity of the system, e.g.~when it possesses a large number of coupled degrees of freedom. By virtue of the fluctuation-dissipation theorem, the non-Markovianity is assisted by the emergence of correlated thermal fluctuations \cite{Kubo1966}, but it can also result from nonequilibrium colored noise \cite{Hanggi1995, Luczka2005}. 
Although the non-Markovian character of the analyzed setups is often neglected, it should be seen as a universal feature, rather than an exception \cite{Kampen1998}.
Even in the simplest systems in quantum mechanics the memoryless assumption leads to the divergence of their energy \cite{Spiechowicz2021}.

A paradigmatic model of a non-Markovian setup is a Brownian particle subjected to power-law correlated thermal fluctuations.
The dynamics of such a system can be described by an integro-differential Generalized Langevin Equation (GLE) \cite{Bogoliubov1945, Ford1965, Kubo1966, Zwanzig1973, Grigolini1982, Hanggi1997}, in which the memory is represented by an appropriate damping kernel.
Analytical solutions of this equation are known for a free particle and a particle in a quadratic potential \cite{Luczka2005, Kupferman2004}, however, in other cases it is not attainable and the analysis must rely on numerical simulations. The latter are problematic unless the memory kernel describes the exponential decay. 

Nowadays, the standard approach to a non-Markovian evolution is known as Markovian embedding, in which it is represented as a projection of a higher dimensional Markovian dynamics \cite{Wang1945, Mori1965, Berne1970, Hanggi1977, Grabert1977, Ferrario1979}. 
To the best of our knowledge, the first implementation of this method dates back to the early 1980s when the Kramers theory was applied to colored noise \cite{Marchesoni1983, Straub1986}. Later it has been employed to study numerous problems, including anomalous diffusion \cite{Goychuk2012, Siegle2010a, Siegle2010b, Goychuk2019}, negative mobility \cite{Kostur2009}, transport in periodic systems \cite{Machura2010}, stochastic thermodynamics \cite{Strasberg2017}, heat engines \cite{Restrepo2018} as well as system-environment correlations \cite{Campbell2018} and machine learning non-Markovian evolution in quantum dynamics \cite{Luchnikov2020}, to mention only a few. In this approach, the original memory kernel is approximated by a finite sum of exponentials \cite{Kupferman2004}. Then the central point is the propagation of multidimensional embedded Markovian dynamics in time with known algorithms. 

Another recently developed technique known as the effective mass approach \cite{Wisniewski2024pre, Wisniewski2024entropy, Wisniewski2024chaos} relies on transforming the original non-Markovian GLE into a much simpler Markovian Langevin equation with a modified mass of the particle.
In this method, the memory effects are reflected solely as a mass correction in the memoryless equation.
This approach can be applied only when the memory time is short, but in return it offers a significant advantage of simplicity. 


Our goal is to compare these two approaches in terms of accuracy and computational performance.
For this purpose we analyze the non-Markovian dynamics of a free Brownian particle with a power law memory kernel for which the two above methods are approximate. We present conditions under which these approaches are correct and provide a scheme for setting free parameters of Markovian embedding to achieve optimal accuracy. 

In section \ref{sec:model}, we present the reference model of a free Brownian particle considered in this work and the approaches of interest to its dynamics.
In the next section, we present methods of analysis together with quantities measuring the accuracy of the employed schemes as well as their computational efficiency.
Then, in section \ref{sec:results}, we compare the approaches and present our main results.
Finally, section \ref{sec:conclusions} summarizes our work.

\section{Model} \label{sec:model}

Since we want to discuss the advantages and drawbacks of the Markovian embedding and effective mass approach to non-Markovian dynamics, in particular to test their precision, we consider the simplest model of a system with memory which can be solved analytically \cite{Kupferman2004}.
It is a free Brownian particle immersed in a medium exhibiting correlated thermal fluctuations.
The dynamics of such a system can be described with a Generalized Langevin Equation (GLE), which reads
\begin{equation} \label{eq:GLE}
    m\dot{v}(t) + \gamma\!\int_0^t\! K(t-u)v(u)\D u = \eta(t),
\end{equation}
where $m$ is the particle mass, $v$ represents its velocity, and $\gamma$ is the coupling constant (friction coefficient) between the system and its environment. The interaction with the medium results in thermal noise $\eta(t)$ representing a stationary Gaussian process of vanishing mean value. 
The damping or dissipation kernel $K(t)$ \cite{Spiechowicz2018pra} is characterized by the memory time $\tau_c$, which, by virtue of the fluctuation-dissipation theorem \cite{Kubo1966}, describes also the correlation time of thermal noise $\eta(t)$
\begin{equation}
    \langle \eta(t) \eta(u) \rangle = \gamma k_B T K(|t-u|).
\end{equation}

We first rescale Eq.~(\ref{eq:GLE}) to a dimensionless form. For this purpose we choose the time, velocity, and position units as follows
\begin{equation}
    t_0 = \frac{m}{\gamma},\qquad v_0 = \sqrt{\frac{k_B T}{m}},\qquad x_0 = v_0t_0,
\end{equation}
where $t_0$ is the so-called Langevin time scale representing the characteristic time of the particle velocity relaxation, and $v_0$ is the square root of the thermal velocity.
Then, we define the new dimensionless variables as
\begin{equation}
    \hat{t} = \frac{t}{t_0},\qquad \hat{v}(\hat{t}) = \frac{v(t)}{v_0},\qquad \hat{x}(\hat{t}) = \frac{x(t)}{x_0}.
\end{equation}
For such a choice of the scaling units, the dimensionless GLE reads
\begin{equation} \label{eq:gle}
    \dot{\hat{v}}(\hat{t}) + \int_0^{\hat{t}} \hat{K}(\hat{t}-\hat{u})\hat{v}(\hat{u})\D \hat{u} = \hat{\eta}(\hat{t}),
\end{equation}
where $\hat{K}(\hat{t}) = K(t)t_0$ is related to $\hat{\eta}(\hat{t})$ by the relation ${\langle \hat{\eta}(\hat{t}) \hat{\eta}(\hat{u}) \rangle = \hat{K}(|\hat{t}-\hat{u}|)}.$ 
To simplify the notation, we omit the hat over the rescaled variables, keeping in mind that we will work with dimensionless quantities from now on.



\subsection{White noise approximation}

If the correlation time $\tau_c$ of thermal fluctuations is much shorter than other characteristic times of the system, the memory effects might be negligible \cite{Wisniewski2024pre} and the GLE (Eq.~\ref{eq:gle}) may be approximated by a much simpler Langevin equation
\begin{equation} \label{eq:white}
    \dot{v}(t) + v(t) = \xi(t),
\end{equation}
where $\xi(t)$ represents white thermal noise obeying $\langle \xi(t) \xi(u) \rangle = 2\delta(t-u)$. The corresponding memory kernel is therefore given by 
\begin{equation} \label{eq:Kw}
    K_\text{w}(t) = 2\delta(t).
\end{equation}
In this approach all non-Markovian effects are neglected and the system can be analyzed with tools developed for a standard Langevin equation. Below we present two methods that allow to take into account the impact of memory.

\subsection{Effective mass approach}

Although the Markovian approximation seems well-justified when the correlation time of thermal fluctuations is much smaller than other time scales, it turns out that even extremely short memory can induce significant differences in system dynamics \cite{Wisniewski2024chaos}.
A very appealing method for capturing these changes in the memoryless equation was recently proposed in \cite{Wisniewski2024pre} and is called the effective mass approach. 

In this method the memory effects are reflected solely in the correction to the particle mass. In particular, the original dimensionless Eq.~(\ref{eq:gle}) is simplified to the memoryless one
\begin{equation} \label{eq:eff_mass}
   m^*\dot{v}(t) + v(t) = \xi(t),
\end{equation}
where $\xi(t)$ as before represents white thermal noise obeying $\langle \xi(t) \xi(u) \rangle = 2\delta(t-u)$, and 
\begin{equation} \label{eq:correction}
	m^* = 1 - \Delta m
\end{equation}
is the effective mass of the particle. The correction $\Delta m$ depends solely on the form of the original memory kernel $K(t)$, i.e.
\begin{equation} \label{eq:dm}
    \Delta m = \int_0^\infty tK(t)\D t.
\end{equation}

Once the mass correction is known, from a mathematical point of view this method is as simple as the standard Markovian approximation, but its accuracy can be much better. 
The effective mass approach can be applied whenever the integral Eq.~(\ref{eq:dm}) exists and the memory time $\tau$ is significantly smaller than the Langevin time (i.e. $\tau \ll 1$ in our dimensionless units). However, we note that in this method the system of interest in the long time limit tends to an asymptotic state which is different but close to the original one since for short memory time $\tau \ll 1$ the mass correction $\Delta m$ in Eq. (\ref{eq:correction}) is small. While doing so features of the archetypal one are retained, e.g. if the asymptotic state of the original system is an equilibrium state then in the effective mass approach it is also an equilibrium state.


\subsection{Markovian embedding}

It is widely known that any form of the GLE can be obtained from the Hamiltonian dynamics of a particle coupled to a thermal bath consisting of harmonic oscillators \cite{Hanggi1997}. Given the fact that the underlying Hamiltonian dynamics is Markovian, it is natural to pursue an approach based on embedding the system of interest to an enlarged phase space so that the extended one obeys a time-local (Markovian) equation. For a thermal bath considered in thermodynamic limit it formally has infinite dimension. The quest is to find an embedding with a minimal number $N$ of additional phase variables so that together with the pair $\{x,v\}$ it forms a Markovian process \cite{Kupferman2004,Siegle2010a}. We note here that even though the multidimensional process is Markovian, the reduced dynamics of $\{x,v\}$ is still non-Markovian \cite{Luczka2005}. It must be contrasted with the effective mass approach where we exploit the memoryless Eq. (\ref{eq:eff_mass}) for which the dynamics of $\{x,v\}$ is Markovian.

The Markovian embedding method is mathematically exact (i.e., it recovers the original GLE) only when the memory kernel represents the exponential decay \cite{Kupferman2004,Siegle2010a}. In particular, for a sum of $N$ exponents
\begin{equation} \label{eq:Kexp}
    K_\text{exp}(t) = \sum\limits_{i=0}^{N-1} c_i e^{-t/\tau_i}, 
\end{equation}
the GLE can be recasted into a Markovian set of $N+2$ equations by introducing $N$ variables $z_i(t)$ \cite{Straub1986}
\begin{equation} \label{eq:sumexp}
\begin{cases}
    \dot{x}(t) = v(t), \\
    \dot{v}(t) = \sum\limits_{i=0}^{N-1} z_i(t), \\
    \dot{z}_i(t) = -\frac{1}{\tau_i} z_i(t) - c_i v(t) + \sqrt{\frac{c_i}{\tau_i}}\xi_i(t),
\end{cases}
\end{equation}
where $\xi_i(t)$ are independent white noise terms obeying $\langle \xi_i(t) \xi_j(u) \rangle = 2\delta(t-u)\delta_{ij}$. This equivalent form of the GLE can be analyzed by using standard methods developed for Markovian processes.

Other memory kernels can be approximated by $K_\mathrm{exp}(t)$ via tuning of $2N+1$ free parameters: $N$, $\tau_i$ and $c_i$; so that they are tractable to Markovian embbeding. E.g. for an algebraically decaying memory kernel we rely on the following representation of the power law function
\begin{equation} \label{eq:xalpha}
    x^{-\alpha} = \frac{1}{\Gamma(\alpha)}\int_0^\infty \frac{1}{u^{\alpha+1}} e^{-x/u} \D u,
\end{equation}
where $\Gamma(\alpha)$ is the (complete) gamma function \cite{Abramowitz1964}.
We can then approximate the integral in Eq.~(\ref{eq:xalpha}) with a finite sum to get Eq.~(\ref{eq:Kexp}).
However it is important to note that in such a case the Markovian embedding procedure is no longer exact, i.e.~the original GLE is recovered only approximately. Therefore the question on the impact of finite summation effects on the so obtained results always arises.

\subsection{Reference memory kernel}

In the following part of the article we would like to compare the effective mass approach and Markovian embedding method in terms of their accuracy and computational performance, as they are basic tools for the analysis of non-Markovian systems. Since the former is always only an approximation, we choose the reference memory kernel in an algebraic form
\begin{equation} \label{eq:Kref}
    K_\text{ref}(t) = \frac{2\tau^2}{\left(t + \tau\right)^{3}},
\end{equation}
for which the Markovian embedding is also not exact. Here $\tau$ stands for the memory time or equivalently the correlation time of thermal fluctuations.

For this kernel the integral in Eq.~(\ref{eq:dm}) exists, so the effective mass approach can be applied.
The mass correction then reads
\begin{equation}
    \Delta m_\mathrm{ref} = \tau.
\end{equation}
Moreover, for the rest of the work we fix ${\tau = 0.01}$, so that it stays within the range of applicability (${\tau \ll 1}$) of the effective mass approach.

As detailed in the previous subsection, for such a reference memory kernel it is not possible to readily exploit the Markovian embedding procedure. However, the kernel can be approximated with a sum of exponents. Here we follow the procedure for choosing the parameters $\tau_i$ presented in \cite{Goychuk2009}, i.e.
\begin{equation}
    \tau_i = \tau_0 b^i,
\end{equation}
where $\tau_0$ and $b$ are constants.
This method of choosing the parameters was originally designed to approximate kernels of type $1/\tau^\alpha$ where $\alpha\in(0, 1)$, but here we incorporate it for $\alpha > 2$.
Moreover, we choose
\begin{equation}
    c_i = \frac{1}{C\tau_i^\alpha} e^{-\tau/\tau_i},
\end{equation}
where $\alpha$ is the exponent of the reference power-law kernel (in our case $\alpha=3$), and 
\begin{equation}
    C = \sum\limits_{i=0}^{N-1} \frac{1}{\tau_i^{\alpha-1}} e^{-\tau/\tau_i}
\end{equation}
is the normalization constant (${\int_0^\infty K_\text{exp}(t) \D t = 1}$).
The $e^{-\tau/\tau_i}$ term follows from the shift of the time variable in Eq.~(\ref{eq:Kref}).


\section{Methods} \label{sec:methods}
In this section we present the quantities used for validation and benchmarking of the effective mass approach and Markovian embedding technique.

\subsection{Accuracy of approximations}

To compare the accuracy of the approximations, we define a difference function
\begin{equation} \label{eq:sigma}
    \sigma = \int_0^\infty |\rho(t)-\rho_\text{ref}(t)| \mathrm{d}t,
\end{equation}
where $\rho_\mathrm{ref}(t)$ is the velocity autocorrelation function (VACF) of the particle in the reference model, whereas $\rho(t)$ represents VACF in one of the approximate systems.
The velocity autocorrelation function is defined as
\begin{equation} \label{eq:vacf}
    \rho(t) = \langle v(t) v(0) \rangle.
\end{equation}
The difference function $\sigma$ measures how much the approximate VACF differs from the reference one.
For the studied model of a free Brownian particle immersed in a correlated thermal bath, the VACF can be expressed in the Laplace space as \cite{Kubo1966, Siegle2011}
\begin{equation} \label{eq:vacf_s}
    \tilde{\rho}(s) = \frac{1}{m^*s + \tilde{K}(s)},
\end{equation}
where $m^* = 1 - \Delta m$ in the effective mass approach and $m^* = 1$ in other schemes.

For the white noise approximation, the unilateral Laplace transform of the kernel (\ref{eq:Kw}) reads $\tilde{K}_\mathrm{w}(s) = 1$ \cite{AL-Jaber2019}, so the VACF can be easily transformed into the time domain and reads
\begin{equation}
    \rho_\mathrm{w}(t) = e^{-t}.
\end{equation}
In the effective mass approach, the memory kernel is the same, but the mass correction needs to be taken into account, i.e.,
\begin{equation}
    \rho_\mathrm{em}(t) = \frac{1}{1-\Delta m} e^{-t/(1-\Delta m)}.
\end{equation}
The sum of exponents in the Laplace space reads
\begin{equation}
    \tilde{K}_\mathrm{exp}(s) = \sum\limits_{i=0}^{N-1} \frac{c_i\tau_i}{s\tau_i + 1},
\end{equation}
whereas the reference kernel transforms to
\begin{equation}
    \tilde{K}_\mathrm{ref}(s) = e^{s\tau}\Gamma(0, s\tau)(s\tau)^2 - s\tau + 1,
\end{equation}
where $\Gamma(0, s\tau)$ is the upper incomplete gamma function \cite{Abramowitz1964}.
For the above kernels $\tilde{K}_\mathrm{exp}(s)$ and $\tilde{K}_\mathrm{ref}(s)$, it is not possible to obtain the VACF in the time domain in a closed form. Therefore we numerically perform the inverse Laplace transform of Eq.~(\ref{eq:vacf_s}) using the Talbot algorithm, in which the VACF in the time domain is approximated by a linear combination of its transform values \cite{Abate2006}. Then the integral in Eq.~(\ref{eq:sigma}) is calculated numerically using the Simpson method for $t$ logarithmically spaced in the range $[10^{-6}, 10^{2}]$.

\subsection{Performance of approximations}

For each of the equations (\ref{eq:white}), (\ref{eq:eff_mass}) and (\ref{eq:sumexp}), we analyzed the number of primitive operations (PO), such as additions, multiplications, assignments, etc., required to implement the weak second-order predictor-corrector algorithm \cite{Platen}, commonly employed in solving stochastic differential equations (see Appendix~\ref{app:pos}).
Assuming that each of these POs takes the same amount of time, we defined the time cost of an approximation as the number of POs per time step required to implement the algorithm.

\section{Results} \label{sec:results}

In general, the essence of the Markovian embedding method lies in the fact that the memory kernel in the original GLE (in our case algebraically decaying $K_\mathrm{ref}(t)$) is approximated with a different one (here $K_\mathrm{exp}(t)$), which can be readily treated numerically. As a consequence, the starting non-Markovian GLE is replaced with another non-Markovian one. In contrast, in the effective mass approach the approximate equation is memoryless (Markovian) and the memory effects are reflected solely in the particle mass renormalization. Therefore this approach is significantly simpler. For this reason one could expect that the Markovian embedding method offers superior accuracy since the memory is directly involved in the approximation. However, we now prove that it is not always true.

In the following subsection we show that the accuracy of Markovian embedding method depends on the fit of the approximate kernel to the reference one and present cases when this method is either more or less accurate than the effective mass approach. In doing so we also demonstrate that the fitting procedure can be nontrivial as common quantifiers characterizing its goodness may lead to suboptimal accuracy.
Then, in the next subsection we provide new algorithm for choosing the correct parameters describing the approximate kernel.
Finally, we compare the computational cost associated with both approximation schemes.

\subsection{Markovian embedding vs the effective mass}

The Markovian embedding method in the studied case requires fixing of three parameters: $b$, $\tau_0$, and $N$. The choice of their values determines the approximate kernel fit quality what is crucial for the precision of this scheme. Typically, the fit quality is quantified by the so-called loss function, which can be defined in several ways. The most straightforward one reads
\begin{equation} \label{eq:I1}
    I_1 = \int_0^\infty |K_\mathrm{ref}(t) - K_\mathrm{exp}(t)|\mathrm{d}t.
\end{equation}
The absolute value in the above formula usually makes analytical evaluation of the integral impossible and the loss function $I_1$ must be calculated numerically.
Another natural choice of the integrand is the squared difference between the kernels, for which the measure reads
\begin{equation} \label{eq:I2}
    I_2 = \int_0^\infty \left(K_\mathrm{ref}(t) - K_\mathrm{exp}(t)\right)^2\mathrm{d}t.
\end{equation}
Finding the parameters of $K_\mathrm{exp}(t)$ by minimizing $I_2$ is the continuous analogue of the least squares method, which is commonly used in e.g.~data fitting. Although these two quantifiers are the standard ones, in the following part of this article we show that they are not suitable for fitting the approximate kernel to achieve the optimal accuracy.

\begin{figure}[t]
    \centering
    \includegraphics[width=\linewidth]{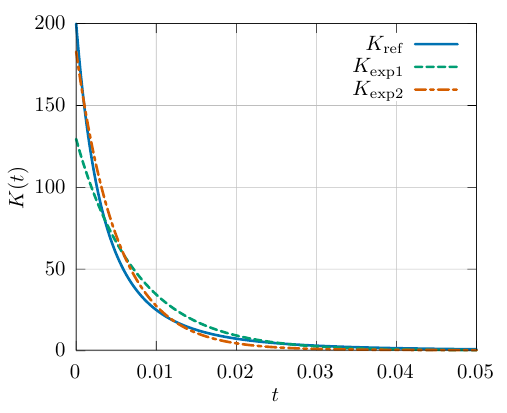}
    \caption{The reference kernel $K_\mathrm{ref}(t)$ plotted along with two approximate kernels $K_\mathrm{exp}(t)$ for $N=10$, $b=10$, and: $\tau_0=7.54\times10^{-4}$ ($K_\mathrm{exp1}$); $\tau_0=5.2\times10^{-4}$ ($K_\mathrm{exp2}$).}
    \label{fig:kernels}
\end{figure}

In Fig.~\ref{fig:kernels} we compare the reference kernel to the approximate ones for two different parameter sets.
Visual inspection of the plots suggests that the fit quality is better for $K_\mathrm{exp2}(t)$, and worse for $K_\mathrm{exp1}(t)$.
This intuitive prediction is confirmed by the values of the quantifiers $I_1$ and $I_2$ calculated for these two parameter sets -- both loss functions take lower values for $K_\mathrm{exp2}(t)$ ($I_1 = 0.917$ and $I_2 = 0.155$) and higher for $K_\mathrm{exp1}(t)$ ($I_1 = 4.505$ and $I_2 = 0.228$). One could thus expect that to achieve the optimal accuracy one should approximate the reference kernel $K_\mathrm{ref}(t)$ with $K_\mathrm{exp2}(t)$. In fact, this is not true. 

\begin{figure}[t]
    \centering
    \includegraphics{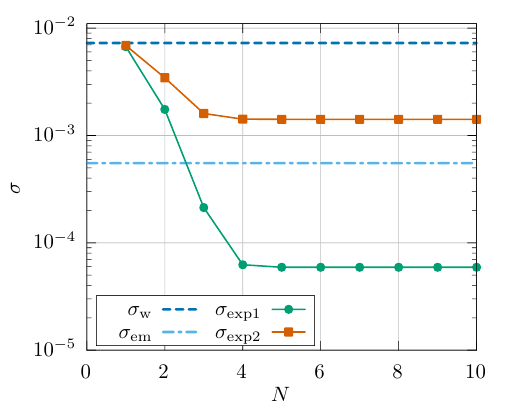}
    \caption{The difference $\sigma$ as a function of the number $N$ of exponents in the memory kernel $K_\mathrm{exp}(t)$ for the same set of parameters as in Fig.~\ref{fig:kernels}. The values for the white noise approximation $\sigma_\mathrm{w}$ and effective mass approach $\sigma_\mathrm{em}$ were marked for reference.}
    \label{fig:sigma_N}
\end{figure}

In Fig.~\ref{fig:sigma_N} we show the difference $\sigma$ measuring the accuracy of the Markovian embedding method as a function of the number $N$ of exponents in the memory kernel $K_\mathrm{exp}(t)$ for the same sets of parameters $b$ and $\tau_0$ as in Fig.~\ref{fig:kernels}.
Surprisingly, the difference $\sigma$ is the lowest for the first set of parameters corresponding to $K_\mathrm{exp1}(t)$. Moreover, for this set the Markovian embedding method is more accurate than the effective mass approach for $N>2$, i.e., $\sigma_\mathrm{exp1} < \sigma_\mathrm{em}$. For the second set of parameters corresponding to $K_\mathrm{exp2}(t)$, for which $I_1$ and $I_2$ are lower, $\sigma_\mathrm{exp2} > \sigma_\mathrm{exp1}$, but also $\sigma_\mathrm{exp2} > \sigma_\mathrm{em}$, regardless of the number $N$ of exponents. This means that for the parameter set chosen by minimizing $I_1$ or $I_2$ the Markovian embedding method performs worse than the effective mass approach. The above discussion shows that the impact of the approximate kernel fitting on the method precision may be nontrivial and therefore a deeper analysis of this problem is needed.

\begin{figure}[t]
    \centering
    \includegraphics[width=\linewidth]{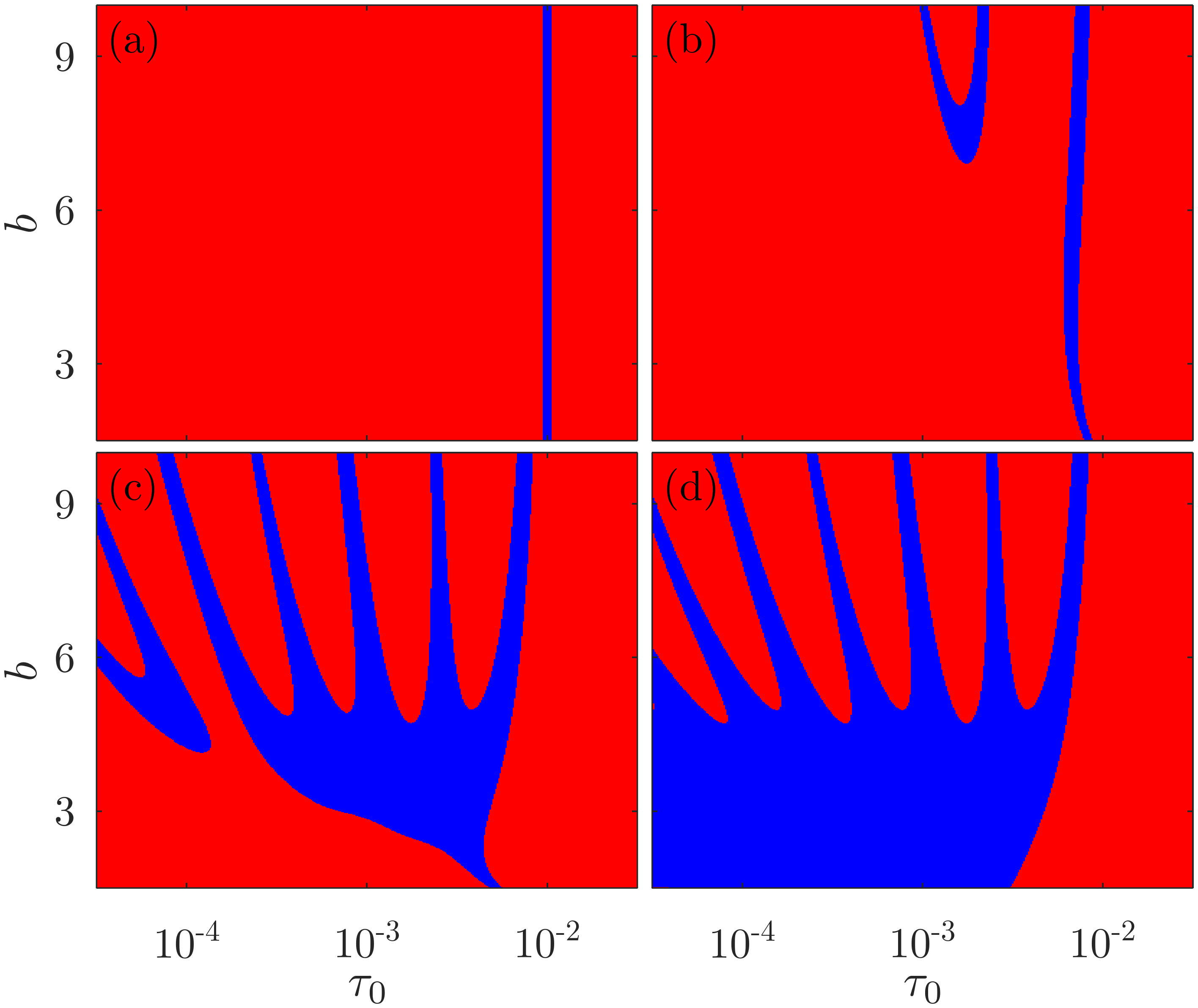}
    \caption{The areas in $(b,\tau_0)$ parameter space, in which the difference $\sigma_\mathrm{exp}$ is greater (red, {\color{red}$\blacksquare$}) or lower (blue, {\color{blue}$\blacksquare$}) than $\sigma_\mathrm{em}$, for selected number $N$ of exponents in $K_\mathrm{exp}(t)$: (a) $N=1$, (b) $N=2$, (c) $N=5$ and (d) $N=20$. For $N=1$ the kernel $K_\mathrm{exp}(t)$ does not depend on $b$ due to normalization.}
    \label{fig:sigma_binary}
\end{figure}

The examples presented above represent two classes of the Markovian embedding method accuracy. Since as we demonstrated the approximation quality depends on the parameters describing the kernel, in Fig.~\ref{fig:sigma_binary} we present a broader, systematic picture, where the $(b, \tau_0)$ parameter space is divided into regions where $\sigma_\mathrm{exp}$ is lower (blue) or higher (red) than $\sigma_\mathrm{em}$. \emph{Mutatis mutandis} the red area corresponds to the situation when the effective mass is more accurate than the Markovian embedding, whereas the blue one indicates the reversed scenario.
When the number $N$ of exponents in the approximation $K_\mathrm{exp}(t)$ is small, it is very difficult to achieve accuracy better than in the effective mass approach with the Markovian embedding method, see panel (a) and (b). When $N$ grows, the blue area corresponding to the situation when the embedding is superior to the effective mass is larger, however, the optimal choice of $b$ and $\tau_0$ is still not obvious and requires a tuning, see panel (c) and (d). 
Moreover, a larger number of exponents results in a more complex model and higher computational cost.

Since Markovian embedding method with correctly chosen parameters can be more precise than the effective mass approach, the natural question arises how to find the parameters $b$ and $\tau_0$ offering the best accuracy for a given $N$? In the next subsection we show that the information from the effective mass approach can serve as an input to Markovian embedding, which allows to achieve the best results with this latter method.

\subsection{Optimal embedding via the effective mass}

The examples presented in the previous subsection showed that the standard loss functions typically used in determining the similarity between the kernels are not suitable for finding parameters for which the difference $\sigma$ is minimized. Moreover, even visual evaluation of the fit quality can be misleading, see Fig.~\ref{fig:kernels}. Here we propose another form of the loss function, which is based on the correction appearing in the definition of the effective mass.
It consists of an absolute value of an integral of the difference between the kernels multiplied by time, i.e.
\begin{equation} \label{eq:Iem}
    I_\mathrm{em} = \left| \int_0^\infty t\left(K_\mathrm{ref}(t) - K_\mathrm{exp}(t)\right)\mathrm{d}t \right|.
\end{equation}
The additional presence of time increases the significance of the long tail of the reference kernel in the fitting process.
A careful comparison of Eqs. (\ref{eq:Iem}) and (\ref{eq:dm}) reveals that this version of the loss function can be written as
\begin{equation}
    I_\mathrm{em} = |\Delta m_\mathrm{ref} - \Delta m_\mathrm{exp}|,
\end{equation}
where $\Delta m_\mathrm{exp}$ is the mass correction corresponding to the approximate kernel $K_\mathrm{exp}(t)$ and reads
\begin{equation} \label{eq:dm_exp}
    \Delta m_\mathrm{exp} = \sum\limits_{i=0}^{N-1} c_i \tau_i^2.
\end{equation}
Consequently, $I_\mathrm{em}$ can be represented in a closed analytical form as
\begin{equation}
	I_\mathrm{em} = \left|\tau - \sum\limits_{i=0}^{N-1} c_i \tau_i^2\right|.
\end{equation}

\begin{figure}[t]
    \centering
    \includegraphics{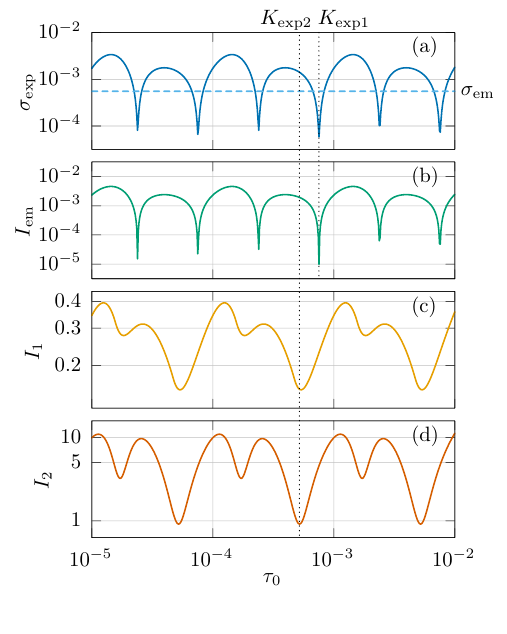}
    \caption{The difference $\sigma_\mathrm{exp}$ (a) and the loss functions $I_\mathrm{em}$ (b), $I_1$ (c) and $I_2$ (d) for $b=10$ and $N=10$ as a function of $\tau_0$. In (a) the value of $\sigma_\mathrm{em}$ was marked for reference. The values of $\tau_0$ corresponding to $K_\mathrm{exp1}(t)$ and $K_\mathrm{exp2}(t)$ were also depicted.}
    \label{fig:comparison}
\end{figure}

To compare the effectiveness of the above defined variants of loss function, in Fig.~\ref{fig:comparison} we present the difference $\sigma_\mathrm{exp}$ together with $I_\mathrm{em}$, $I_1$ and $I_2$ for $b=10$ and $N=10$ as a function of $\tau_0$.
The quantities $I_1$ and $I_2$ do not correlate well with the difference $\sigma_\mathrm{exp}$ and therefore they are not useful for finding the optimal parameters of the embedding guaranteeing the best accuracy. We have confirmed that this observation holds true also for other, not depicted parameter regimes.
On the contrary, in Fig.~\ref{fig:comparison} there is a stunning agreement between minimas of $I_\mathrm{em}$ and those corresponding to high accuracy of the Markovian embedding method in $\sigma_\mathrm{exp}$. 
This means that the discrepancy $I_\mathrm{em} = |\Delta m_\mathrm{ref} - \Delta m_\mathrm{exp}|$ between the mass correction calculated in the effective mass approach can be used as a measure of similarity between the reference and approximate memory kernels. This also explains why the Markovian embedding approximation performs better for $K_\mathrm{exp1}(t)$ (parameters corresponding to a minimum of $I_\mathrm{em}$) than for $K_\mathrm{exp2}(t)$ (corresponding to a minimum of $I_2$ and very close to a minimum of $I_1$). We have checked numerically that the correspondence between $\sigma_\mathrm{exp}$ and $I_\mathrm{em}$ extends even to correlation times $\tau$ beyond the range of applicability of the effective mass approach $\tau \ll 1$.

\begin{figure}[t]
    \centering
    \includegraphics[width=\linewidth]{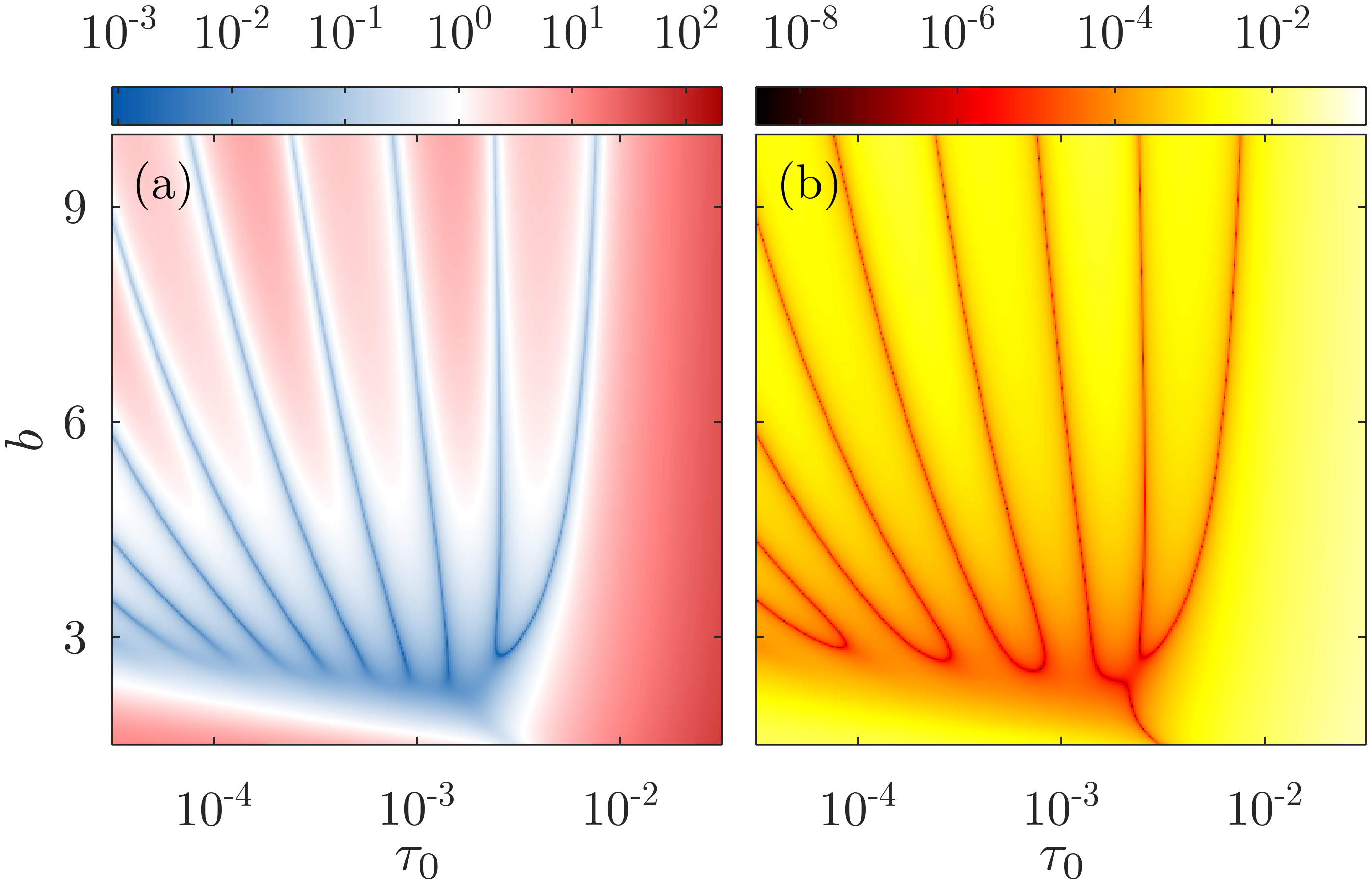}
    \caption{Panel (a): the relative difference $\sigma_\mathrm{exp}/\sigma_\mathrm{em}$. Panel (b): the absolute difference between the mass corrections \mbox{$I_\mathrm{em} = |\Delta m_\mathrm{ref} - \Delta m_\mathrm{exp}|$}. The remaining parameter reads $N = 10$.}
    \label{fig:sigma_dm_map}
\end{figure}
Finally, in Fig.~\ref{fig:sigma_dm_map}a, the difference $\sigma_\mathrm{exp}$ relative to $\sigma_\mathrm{em}$ is plotted as a function of $b$ and $\tau_0$ for $N=10$.
Similarly, Fig.~\ref{fig:sigma_dm_map}b shows the discrepancy $I_\mathrm{em}$ between the mass corrections for the approximate and the reference kernel. The stunning qualitative similarity of these plots confirms that the results for the sum of exponents $K_\mathrm{exp}(t)$ are closest to those obtained for the reference kernel $K_\mathrm{ref}(t)$ when the mass correction $\Delta m_\mathrm{exp}$ related to it matches $\Delta m_\mathrm{ref}$. This, in turn, means that in such a case the parameters $b$ and $\tau_0$ are optimal for a given $N$, and the precision of the approximation is the best.
The only noticeable difference is in the area where $b$ and $\tau_0$ are small.
Careful inspection of Fig.~\ref{fig:sigma_binary} indicates that in this region $\sigma_\mathrm{exp}$ saturates much slower and reaches its final value for a relatively high number of exponents $N$, in this case for $N>10$.
For larger $N$, the agreement between plots \ref{fig:sigma_dm_map}a and \ref{fig:sigma_dm_map}b is extended also to this area.

\subsection{The cost of precision}

With the efficient technique of choosing parameter values for the Markovian embedding method presented in the previous section, one gets a powerful apparatus for approximating Eq.~(\ref{eq:gle}) with kernels $K(t)$ exhibiting different from the exponential decay.
However, this comes with a cost.
Firstly, implementing the numerical algorithm to solve the equations of motion in the Markovian embedding method is much more complicated than in the effective mass approach, cf. Eq.~(\ref{eq:sumexp}) and Eq.~(\ref{eq:eff_mass}).
Secondly, the higher accuracy of the Markovian embedding method typically requires a larger number of exponents, which results in a significant computational cost.
Below, we compare the theoretical performance of all presented approximations.

Table \ref{tab:po} shows the number of primitive operations required to implement a weak second-order predictor-corrector algorithm \cite{Platen} for each of the approximate approaches.
\begin{table}[b]
\caption{\label{tab:po}
Number of primitive operations (POs) per time step required to solve the equations of motion for each approximation.
}
\begin{ruledtabular}
\begin{tabular}{lc}
\multicolumn{1}{c}{Method} & Number of POs \\
\colrule
White noise                & $99$    \\
Effective mass             & $115$    \\
Markovian embedding        & $41+99N$ \\
\end{tabular}
\end{ruledtabular}
\end{table}
The white noise approximation is the least complex since the memory effects are completely neglected, and the integro-differential equation (\ref{eq:gle}) is replaced by a much simpler differential equation (\ref{eq:white}).
In our implementation of the numerical algorithm, this method requires 99 primitive operations (POs) per time step, including the generation of the stochastic component (see Appendix \ref{app:pos} for the detailed analysis).

The effective mass approach is essentially as simple as the white noise approximation since the only difference is the presence of the mass correction in Eq.~\ref{eq:eff_mass}.
Hence, the number of POs per time step is practically the same and totals 115.

\begin{figure}
    \centering
    \includegraphics[width=\linewidth]{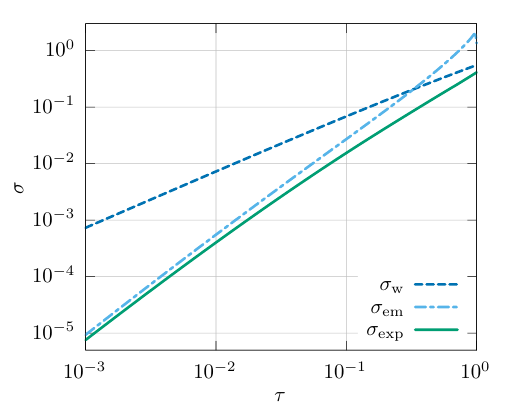}
    \caption{The difference $\sigma$ as a function of the correlation time $\tau$ for the white noise approximation, effective mass approach and Markovian embedding with $N=1$. In the Markovian embedding method the parameter $\tau_0=\tau$, so that $\Delta m_\mathrm{exp} = t\int_0^\infty tK_\mathrm{exp}(t) \mathrm{d}t$.}
    \label{fig:sigma_tau}
\end{figure}
The Markovian embedding method is significantly more complicated than the previous approaches.
The presence of the parameter $N$ makes the time performance of this method dependent on the number of exponents constituting the approximate kernel $K_\mathrm{exp}(t)$.
In our implementation, the number of POs per time step for this method is $41+99N$.
This means that even for $N=1$ the algorithm is slower than the effective mass approach.

Moreover, the accuracy advantage of Markovian embedding for $N=1$ over the effective mass approach is not significant. 
In Fig.~\ref{fig:sigma_tau} we present how the difference $\sigma$ depends on the correlation time $\tau$ for the white noise approximation, the effective mass approach and the Markovian embedding with $N=1$ and the parameter $\tau_0=\tau$ (so that $\Delta m_\mathrm{exp} = t\int_0^\infty tK_\mathrm{exp}(t) \mathrm{d}t$). 
In the whole range of $\tau\in(10^{-3}, 10^{-1})$ the difference $\sigma$ is smaller for the Markovian embedding method, but the difference is minute.

In practice, to obtain the desired accuracy, one needs to set the number of exponents $N$ to a higher value, which implies a significant increase in the computational cost.
For the first set of parameters presented in Fig.~\ref{fig:sigma_N} (corresponding to $\sigma_\mathrm{exp1}$), the Markovian embedding method gives more accurate results than the effective mass approach for $N=3$.
This means that the computational cost of achieving $\sigma_\mathrm{exp1} < 10^{-3}$ is almost three times higher than in the much simpler effective mass method.
Much higher precision can be achieved when both $b$ and $\tau_0$ are small, however then the number of exponents required to approximate the original kernel correctly is even higher so that the Markovian embedding scheme can be an order of magnitude slower than the effective mass approach.
The question arises whether such great accuracy is necessary and if the increased precision have any visible impact on the simulation results.

The above discussion suggests that whenever the accuracy of the effective mass approach is satisfying, it should be preferred over the Markovian embedding method.
If higher precision is needed, the effective mass approach guides for choosing optimal $(b, \tau_0)$ pair for a given $N$.

\section{Conclusions} \label{sec:conclusions}

In this work we considered a vital problem of dynamics of non-Markovian systems formulated in the framework of the Generalized Langevin Equation. In particular, we focused on a comparison of two methods allowing to tackle such setups, namely the Markovian embedding technique and the recently developed effective mass approach, in terms of their accuracy and computational performance. In doing so we considered a paradigmatic model of a Brownian particle subjected to algebraically correlated thermal fluctuations for which the two above methods are approximate.

Firstly, we have shown that when the memory time is much shorter than other time scales of the system, the lately proposed effective mass approach offers much better accuracy (one order of magnitude) than the commonly employed white noise approximation. Moreover, at the same time it requires practically no additional computational cost, and therefore in such a regime it is a preferred method.

Secondly, when the memory time is not short, one needs to rely on the Markovian embedding technique. We have found that the precision of this method is very sensitive to the parameters describing the approximate memory kernel, which is expressed as a finite sum of exponentially decaying functions. If the parameters are not tuned carefully enough, the resultant accuracy of this method may be worse even than the white noise approximation which completely neglects the memory effects. On the other hand, there are cases when the accuracy of the Markovian embedding method is superior than the effective mass approach, however, this typically comes at the cost of significant growth of the computational complexity. 

Thirdly, taking into account the latter fact, a natural question arises whether there is a robust scheme that allows to find the optimal parameters describing the approximate memory kernel offering the best accuracy and minimal computational cost. It turns out that the concept of effective mass can serve for this purpose. Specifically, we have shown that if the mass correction term calculated for the approximate memory kernel (the finite sum of exponents) corresponds to that of the original one, the accuracy of the Markovian embedding method is radically improved. In other words, the mass correction can be regarded as a measure of similarity between memory kernels that allows to find optimal values for parameters of the method to achieve the best accuracy and minimize the computational cost.

Since the role of memory in the dynamics of systems is a problem that seemingly attracts everlasting activity in physics, and our paper provides a blueprint for such investigations, we expect the emergence of vibrant follow-up works on this and related topics.

\section*{Acknowledgment}
This work was supported by the Grant NCN 2022/45/B/ST3/02619 (J.S.)

\appendix

\section{Primitive operations for each approximation} \label{app:pos}

To solve Eqs.~(\ref{eq:white}), (\ref{eq:eff_mass}) and (\ref{eq:sumexp}), we implemented a weak second-order predictor-corrector algorithm \cite{Platen}.
To explain how this method works, let us consider a differential equation
\begin{equation}
    \dot{\mathbf{X}}(t) = \mathbf{G}(\mathbf{X}(t), t) + \boldsymbol{\xi}(t),
\end{equation}
where $\mathbf{X}$ is a vector of the variables of interest, $\mathbf{G}(\mathbf{X}(t), t)$ is a vector of arbitrary functions of $\mathbf{X}$ and $t$, $\boldsymbol{\xi}(t)$ is a vector of noise components, and the dot represents differentiation with respect to the variable $t$.
Let us now divide the time domain into equal segments of width $\Delta t$ and denote $t_k = k\Delta t$, $\mathbf{X}_k = \mathbf{X}(t_k)$.
The algorithm then works in three steps:
\begin{enumerate}[leftmargin=*]
    \item $\mathbf{X}^P_{k+1}\! =\! \mathbf{X}_k + \mathbf{G}(\mathbf{X}_k, t_k) \Delta t + \boldsymbol{\xi}(t_k)$,
    \item $\mathbf{X}^C_{k+1}\! =\! \mathbf{X}_k + \frac{1}{2}\left[\mathbf{G}(\mathbf{X}_k, t_k) + \mathbf{G}(\mathbf{X}^P_{k+1}, t_{k+1})\right]\Delta t {+ \boldsymbol{\xi}(t_k)}$,
    \item $\mathbf{X}_{k+1}\! =\! \mathbf{X}_k + \frac{1}{2}\left[\mathbf{G}(\mathbf{X}_k, t_k) + \mathbf{G}(\mathbf{X}^C_{k+1}, t_{k+1})\right]\Delta t {+ \boldsymbol{\xi}(t_k)}$,
\end{enumerate}
where the first equation is the predictor step, the two next equations are the corrections, and the vector $\mathbf{X}_{k+1}$ contains the estimated values of $\mathbf{X}$ in time ${t_{k+1} = (k+1)\Delta t}$.
The elements of the noise vector $\boldsymbol{\xi}(t_k)$ return an independent value in each step.
The number of primitive operations (POs) required to implement this scheme in each approximation depends on the complexity of the functions in $\mathbf{G}(\mathbf{X}(t), t)$, and the number of equations for which the predictor-corrector method has to be implemented.

\subsection{White noise approximation}

In the white noise approximation, the vector $\mathbf{X}_k$ consists of the position $x_k = x(t_k)$ and the velocity $v_k = v(t_k)$ variables, i.e.
\begin{equation}
    \mathbf{X}_k = [x_k, v_k].
\end{equation}
The function vector $\mathbf{G}(\mathbf{X}_k, t_k)$ reads
\begin{equation}
    \mathbf{G}(\mathbf{X}_k, t_k) = [g_x(\mathbf{X}_k, t_k), g_v(\mathbf{X}_k, t_k)] = [v_k, -v_k].
\end{equation}
The random component appears only in the equation for $v(t)$, therefore the noise vector reads
\begin{equation}
    \boldsymbol{\xi}(t_k) = [0, \xi(t_k)],
\end{equation}
where $\xi(t_k)$ is implemented as a three-state random variable with the following properties
\begin{equation}
	P\left(\xi(t_k) = \pm\sqrt{6\Delta t}\right) = \frac{1}{6},\qquad P\left(\xi(t_k) = 0\right) = \frac{2}{3}.
\end{equation}
Such a three-state process has analogous moment properties to a Gaussian random variable and can be used as a substitute for the latter in numerical simulations \cite{Platen}.
Implementation of $\xi(t_k)$ includes the generation of a pseudorandom number uniformly distributed over the interval $(0,1)$, and returning a proper final value of $\xi(t_k)$.
The whole process reads
\begin{enumerate}
	\item $r = r \oplus (r \gg 13)$,
	\item $r = r \oplus (r \ll 17)$,
	\item $r = r \oplus (r \gg 5)$,
	\item $r = r \gg 32$,
	\item if $r < 1/6$ then return $-\sqrt{6\Delta t}$,
	\item else if $r > 5/6$ then return $\sqrt{6\Delta t}$,
	\item else return $0$,
\end{enumerate}
where $\oplus$ stands for the logical XOR operation, $\ll$ and $\gg$ represent left and right bitshift, and $r$ is a static variable, i.e.~its value does not reset when the function $\xi$ terminates.

Evaluation of $x_{k+1}$ costs 23 primitive operations in each step, i.e., 3 assignments, 5 additions, 5 multiplications, 5 calls of the function $g_x$, and 1 PO per each call of the function $g_x$ (for returning the value).
The algorithm for $v_{k+1}$ is more complicated and requires 3 assignments, 8 additions, 5 multiplications, 5 calls of the functions $g_v$, and 3 calls of the function $\xi$.
The function $g_v$ costs 2 POs per each call (multiplication by $-1$ and returning the value), and the noise function $\xi$ requires 14 POs for each evaluation.
This gives 76 POs per time step for variable $v_{k+1}$.
The total cost of the whole algorithm is then $23+76 = 99$ POs.

\subsection{Effective mass approach}

In the effective mass approach, the equation of motion differs from the one in the white noise approximation only in the modified mass of the particle.
Thus, the function $g_v(\mathbf{X}_k, t_k)$ changes to
\begin{equation}
    g_v(\mathbf{X}_k, t_k) = -v_k / (1-\Delta m),
\end{equation}
and the noise vector reads
\begin{equation}
    \boldsymbol{\xi}(t_k) = [0, \xi(t_k)/(1-\Delta m)],
\end{equation}
where the random variable $\xi$ is defined the same as in the white noise approximation.
In both cases the additional term costs 2 POs per function call, so in total the whole algorithm requires $115$ POs.

\subsection{Markovian embedding}

The Markovian embedding method requires solving $N+2$ differential equations -- one for the position $x$, one for velocity $v$, and $N$ for the auxiliary variables $z_i$.
The variable vector thus reads
\begin{equation}
    \mathbf{X}_k = [x_k, v_k, z_{i,k}].
\end{equation}
The function vector is then
\begin{equation}
    \mathbf{G}(\mathbf{X}_k, t_k) = [g_x(\mathbf{X}_k, t_k), g_v(\mathbf{X}_k, t_k), g_{z_i}(\mathbf{X}_k, t_k)],
\end{equation}
and its elements read
\begin{align}
    g_x(\mathbf{X}_k, t_k) &= v_k, \\
    g_v(\mathbf{X}_k, t_k) &= \sum\limits_{i=0}^{N-1} z_{i,k}, \\
    g_{z_i}(\mathbf{X}_k, t_k) &= -\frac{1}{\tau_i}z_{i,k} - c_i v_k.
\end{align}
The noise vector is as follows
\begin{equation}
    \boldsymbol{\xi}(t_k) = [0, 0, \sqrt{\frac{c_i}{\tau_i}}\xi_i(t_k)],
\end{equation}
where the noise functions $\xi_i$ are defined the same as $\xi$ and are independent on each other.
The cost of evaluation of $x_{k+1}$ does not differ from the previous approaches, thus it costs 23 POs.
The function $g_v$ requires $N-1$ additions and one value return, therefore it costs $N$ primitive operations per call.
Consequently, evaluation of $v_{k+1}$ requires $18+5N$ POs.
The cost of each of the functions $g_{z_i}$ is 5 (3 multiplications, 1 subtraction, and 1 value return), and there is one additional multiplication by a constant per each call of the noise function. 
This means that there are 94 primitive operations per each variable $z_i$, and in total this gives $23 + (18+5N) + 94N = 41 + 99 N$ POs for the whole algorithm.


\end{document}